\documentclass{svjour3}                     
\smartqed  
\usepackage{graphicx}
\newcommand{\pbarp}{{\bar p p}}
\newcommand{\nbarn}{{\bar N  N}}
\newcommand{\lbarl}{{\bar \Lambda \Lambda}}
\newcommand{\lcbarlc}{\bar{\Lambda}_c^- {\Lambda}_c^+}
\newcommand{\ddbar}{\bar{D}{D}}
\newcommand{\kkbar}{\bar{K}{K}}
\begin{document}

\title{Charm production in antiproton-proton annihilation\footnote{Presented at the 21st 
European Conference on Few-Body Problems in Physics, Salamanca, Spain, 30 August - 3 September 2010.
}
}

\author{J. Haidenbauer \and 
        G. Krein 
}

\institute{J. Haidenbauer \at
Institute for Advanced Simulation, 
Forschungszentrum J\"ulich, D-52425 J\"ulich, Germany \\
              \email{j.haidenbauer@fz-juelich.de}           
\and
           G. Krein \at
Instituto de F\'{\i}sica Te\'{o}rica, Universidade Estadual
Paulista, \\
Rua Dr. Bento Teobaldo Ferraz, 271 - 01140-070 S\~{a}o Paulo, SP, Brazil \\
              \email{gkrein@ift.unesp.br}           
}

\date{Received: date / Accepted: date}

\maketitle

\begin{abstract}
We study the production of charmed mesons ($D$) and baryons 
($\Lambda_c$) in antiproton-proton ($\bar pp$) annihilation close 
to their respective
production thresholds. The elementary charm production process
is described by either baryon/meson exchange or by 
quark/gluon dynamics. Effects of the interactions in the
initial and final states are taken into account rigorously. 
The calculations are performed in close analogy to our earlier
study on $\bar pp \to \bar \Lambda \Lambda$ and $\bar pp \to \bar KK$ 
by connecting the processes via SU(4) flavor symmetry. 
Our predictions for the $\bar \Lambda_c \Lambda_c$ 
production cross section are in the
order of 1 to 7 $mb$, i.e. a factor of around 10-70 
smaller than the corresponding cross sections for $\bar \Lambda \Lambda$
However, they are 100 to 1000 times larger than
predictions of other model calculations in the literature.
On the other hand, the resulting cross sections for $\bar DD$ production
are found to be in the order of
$10^{-2}$ -- $10^{-1}$ $\mu b$ and they turned out to be
comparable to those obtained in other studies. 
\end{abstract}

\section{Introduction}
\label{intro}
The study of the production of charmed hadrons in antiproton-proton ($\pbarp$)
collisions is of importance for the understanding of the strong force in the
nonperturbative regime of QCD. The FAIR project at the GSI site has an
extensive program aiming at a high-accuracy spectroscopy of charmed hadrons and at
an investigation of their interactions with ordinary matter \cite{PANDA}.
Presently little is known
about such interactions and their knowledge is a prerequisite for investigating issues
like in-medium properties of charmed hadrons, e.g. $c\bar{c}$-quarkonium dissociation
and changes in properties of $D$ mesons due to chiral symmetry restoration effects
on the light quarks composing these mesons. 

In this contribution we present predictions for the charm-production 
reactions $\pbarp \rightarrow \lcbarlc$ and $\pbarp \rightarrow \bar DD$ close to 
their respective thresholds \cite{HK10,HK11}.
The work builds on the J\"ulich meson-baryon models
for the reactions $\pbarp \rightarrow \lbarl$~\cite{Haidenbauer:1991kt}
and $\pbarp \rightarrow \kkbar$ \cite{Mull}. 
The extension of those models from the strangeness to the charm
sector follows a strategy similar to our recent work on the $DN$ and ${\bar D}N$
interactions~\cite{Haidenbauer:2007jq,Hai10}
namely by assuming as a working hypothesis SU(4) symmetry constraints. 
We also compare our results with those of other model calculations in the literature 
\cite{Kroll:1988cd,Kaidalov:1994mda,Kerbikov,Titov:2008yf,Goritschnig:2009sq}. 
In some of these studies a quark-gluon description based on a factorization 
hypothesis of hard and soft processes \cite{Kroll:1988cd,Goritschnig:2009sq} 
is employed, while in others
a non-perturbative quark-gluon string model is used, based on secondary Regge pole
exchanges including absorptive corrections~\cite{Kaidalov:1994mda,Titov:2008yf}. 

\section{The model}
\label{sec:1}
The calculations of the charm-production reactions $\pbarp \rightarrow \lcbarlc$
and $\pbarp \rightarrow \bar DD$ are done in complete analogy  
to past investigations of the strangeness-production reactions
$\pbarp \rightarrow \lbarl$~\cite{Haidenbauer:1991kt}
and $\pbarp \rightarrow \kkbar$ \cite{Mull} by the J\"ulich group. 
In particular $\pbarp \rightarrow \lcbarlc$ is performed within a coupled-channel 
approach while for $\pbarp \rightarrow \ddbar$ a DWBA approach is employed.
This allows us to take into account rigorously the effects of the initial ($\pbarp$) and 
also of the final state interactions which play an important role for energies near 
the production threshold \cite{Haidenbauer:1991kt,Kohno}.
 
Because of the known sensitivity of the results for the cross sections on 
the initial $\pbarp$ interaction we examine its effect by considering  
several variants of the $\nbarn$ potential. Details of those potentials
can be found in Ref.~\cite{HK10,HK11}. Here we just want to mention that
they differ primarily in the elastic part where we consider variations
from keeping only the longest ranged contribution (one-pion exchange) to 
taking a full G-parity transformed $NN$ interaction as
done in \cite{Haidenbauer:1991kt}. 
All these models reproduce the total $\pbarp$ cross section in the relevant
energy range and, in general, describe also data on
integrated elastic and charge-exchange cross sections and even
$\pbarp$ differential cross sections, cf.~\cite{HK10}.

The microscopic charm-production processes are described by either
meson exchange ($D$, $D^*$) in case of $\pbarp \rightarrow \lcbarlc$,
or baryon exchange ($\Lambda_c$, $\Sigma_c$) for $\pbarp \rightarrow \ddbar$.
The transition potentials are derived from the 
corresponding transitions in the strangeness-production channels
($\lbarl$, $\bar KK$) under the assumption of SU(4) symmetry. 
In order to assess uncertainties in the model we study,
in addition, the effect of replacing the meson-exchange transition potential 
by a charm-production mechanism derived in a quark model \cite{HK10,Kohno}. 

\section{Results for the reaction $\pbarp \rightarrow \lcbarlc$}
\label{sec:2}
\begin{figure}
\includegraphics[height=60mm,angle=-90]{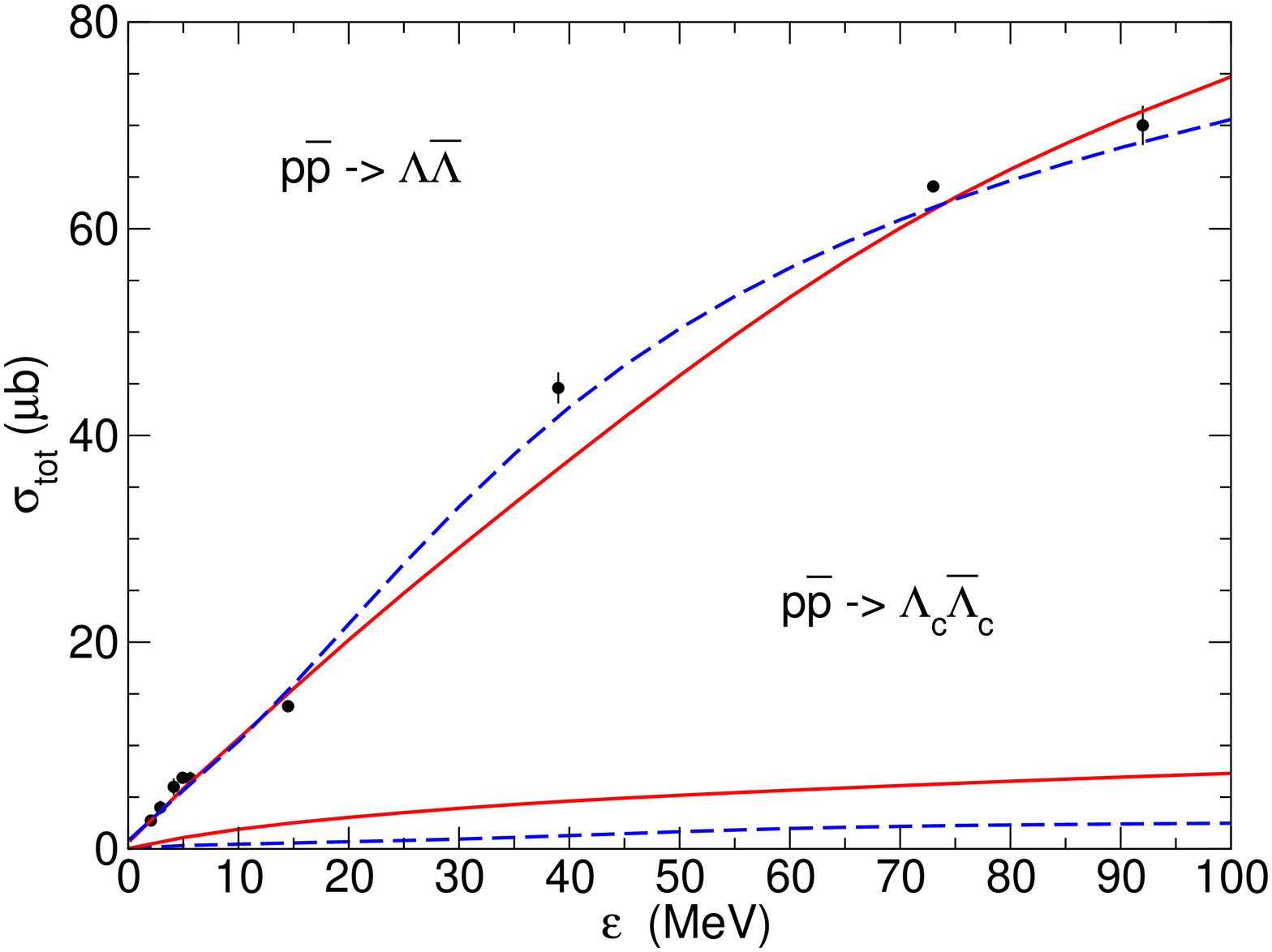}
\includegraphics[height=60mm,angle=-90]{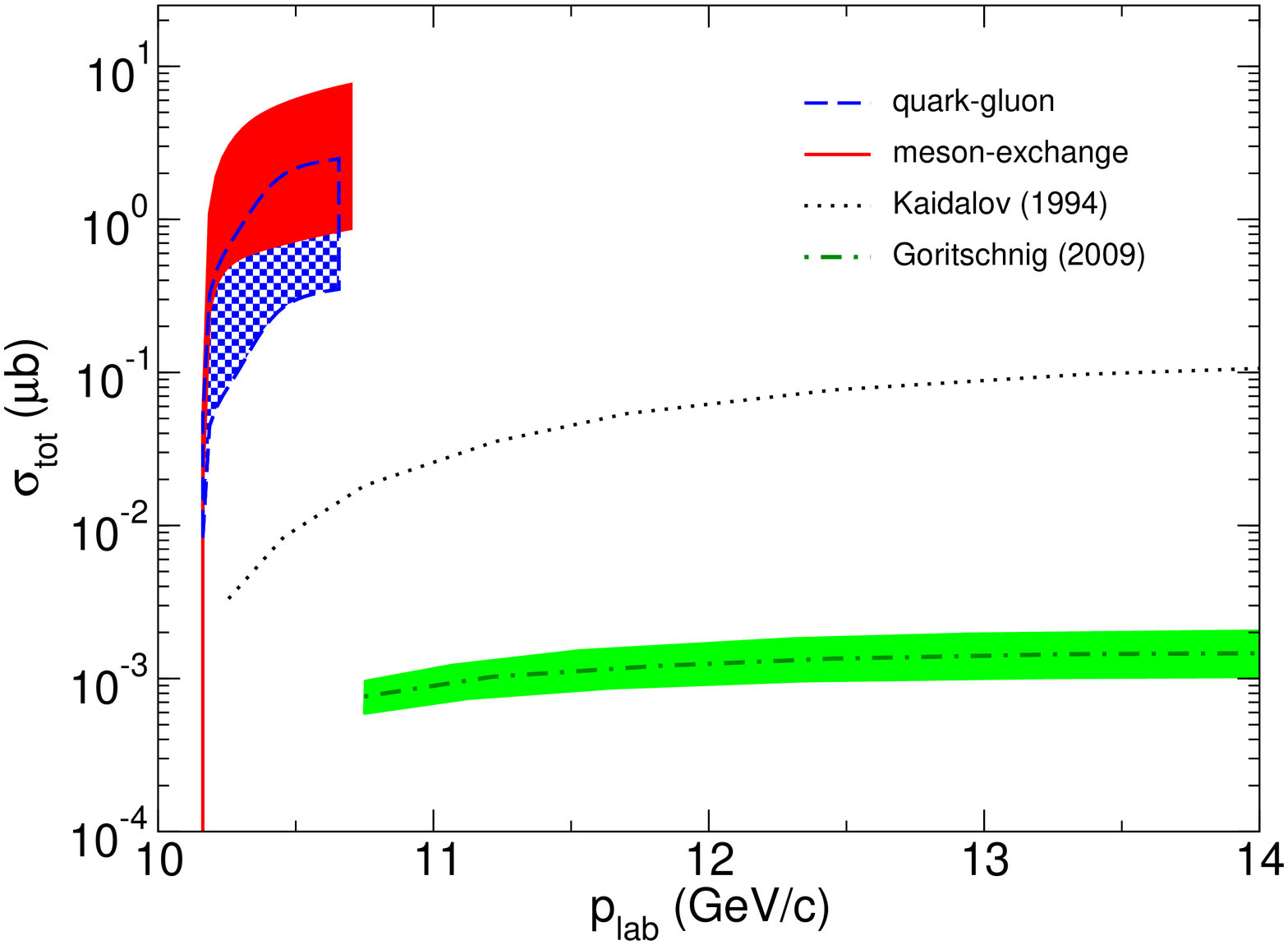}
\caption{Left: Total reaction cross sections for
$\pbarp \to \lbarl$ and $\pbarp \to \lcbarlc$
as a function of the excess energy $\epsilon$.
The solid curves are results for the meson-exchange
transition potential while the dashed curves
correspond to quark-gluon dynamics.
Right: Total reaction cross sections for
$\pbarp \to \lcbarlc$ as a function of $p_{lab}$.
Our predictions (shaded band, grid) are compared to those
by Refs. \cite{Kaidalov:1994mda} and \cite{Goritschnig:2009sq}.
}
\label{fig:2}       
\end{figure}

Predictions for the total reaction cross section
for $p\bar{p} \to \lcbarlc$ are presented
in Fig.~\ref{fig:2}. In the left panel 
the cross section is shown as a function
of the excess energy $\epsilon = \sqrt{s} - m_{\Lambda_c}
- m_{\bar\Lambda_c}$ so that we can compare it with the one
for $p\bar{p} \to \lbarl$ at the corresponding
$\epsilon$. The curve in Fig.~\ref{fig:2} correspond to
the $\pbarp$ model C \cite{HK10}.

Obviously, and as expected, the cross section for $\lcbarlc$
production is smaller than the one for $\lbarl$. But the
difference is about one order of magnitude only.
We display here also the results based on an adaption of the $^3S_1$
quark-gluon transition mechanism of Ref.~\cite{Kohno}.
We scale the effective quark-gluon coupling strength,
fixed in our study of $\pbarp \to \lbarl$~\cite{Haidenbauer:1991kt},
with $ (m_c / m_s)^2$ using the constituent quark masses
$m_s = 550~{\rm MeV}$ and $m_c = 1600~{\rm MeV}$, 
i.e. the same values as employed in our previous works in
Ref.~\cite{Haidenbauer:2007jq}.
As expected, we obtain cross sections that are of the same magnitude as
those predicted in the meson-exchange picture though roughly a factor 
three smaller, cf. the dashed line in Fig.~\ref{fig:2}.

On the right-hand side of Fig.~\ref{fig:2} we compare our predictions with 
those by other groups \cite{Kaidalov:1994mda,Goritschnig:2009sq}. 
Our results are shown as bands in order to reflect the variation of the 
predictions when different ISI's are used. The
dark (red) shaded band and the (blue) grid correspond to results based
on the meson-exchange and quark-gluon transition potentials, respectively. 
It is remarkable that our results differ drastically
from those of the preceeding works. Specifically, our
cross sections are a factor 1000 larger than those given
in \cite{Goritschnig:2009sq} and they are still about 100 times
larger than the ones in \cite{Kaidalov:1994mda}. Thus,
even when considering the variation of about a factor
ten due to the ISI that we see in our results and the
uncertainties due to the unconstrained FSI and form factors
in the transition potential that amount to roughly a factor 
three \cite{HK10}, we are faced with an impressive qualitative 
difference.

\section{Results for the reaction $\pbarp \rightarrow \ddbar$}
\label{sec:3}
\begin{figure}
\includegraphics[height=60mm,angle=-90]{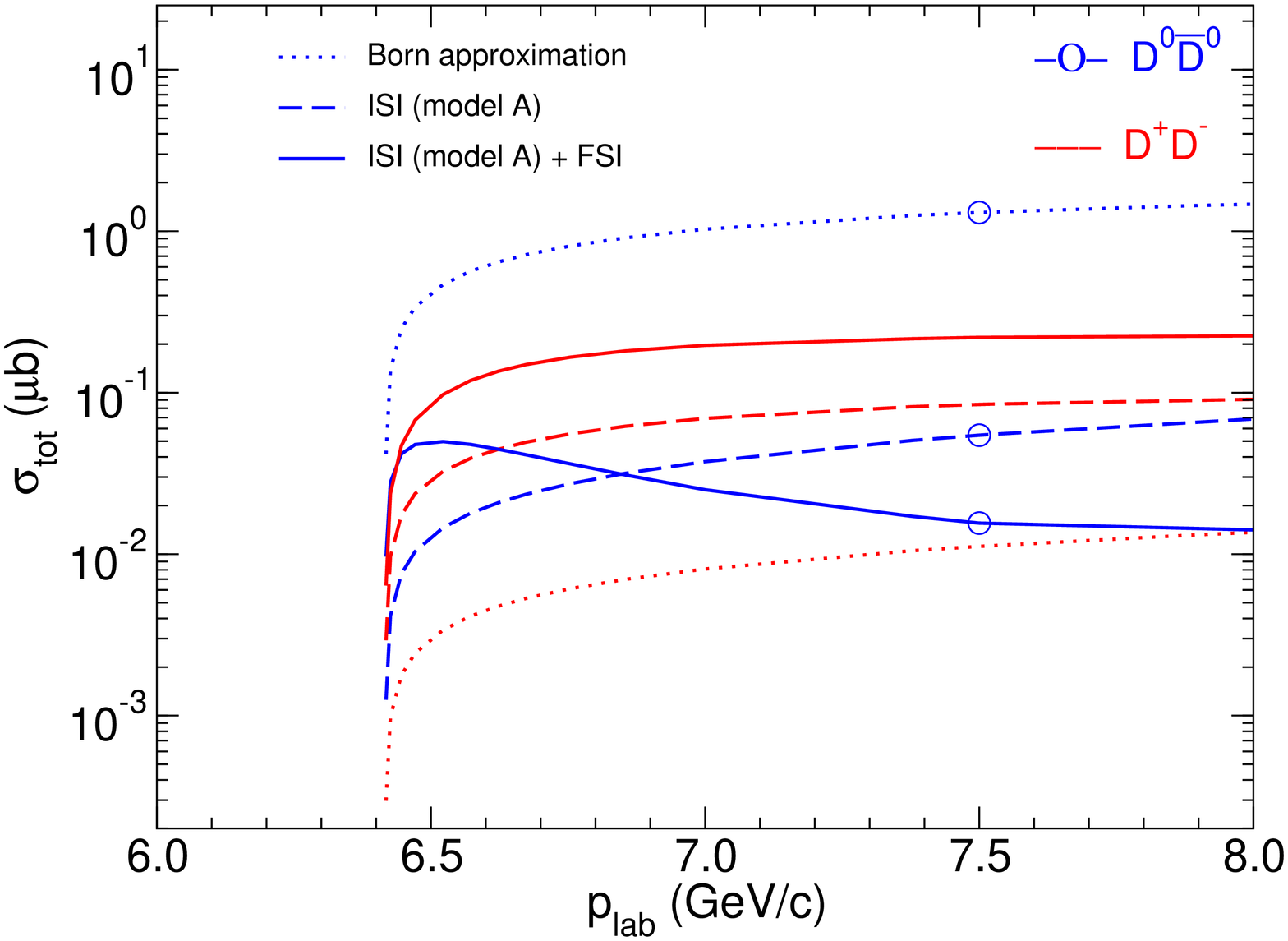}
\includegraphics[height=60mm,angle=-90]{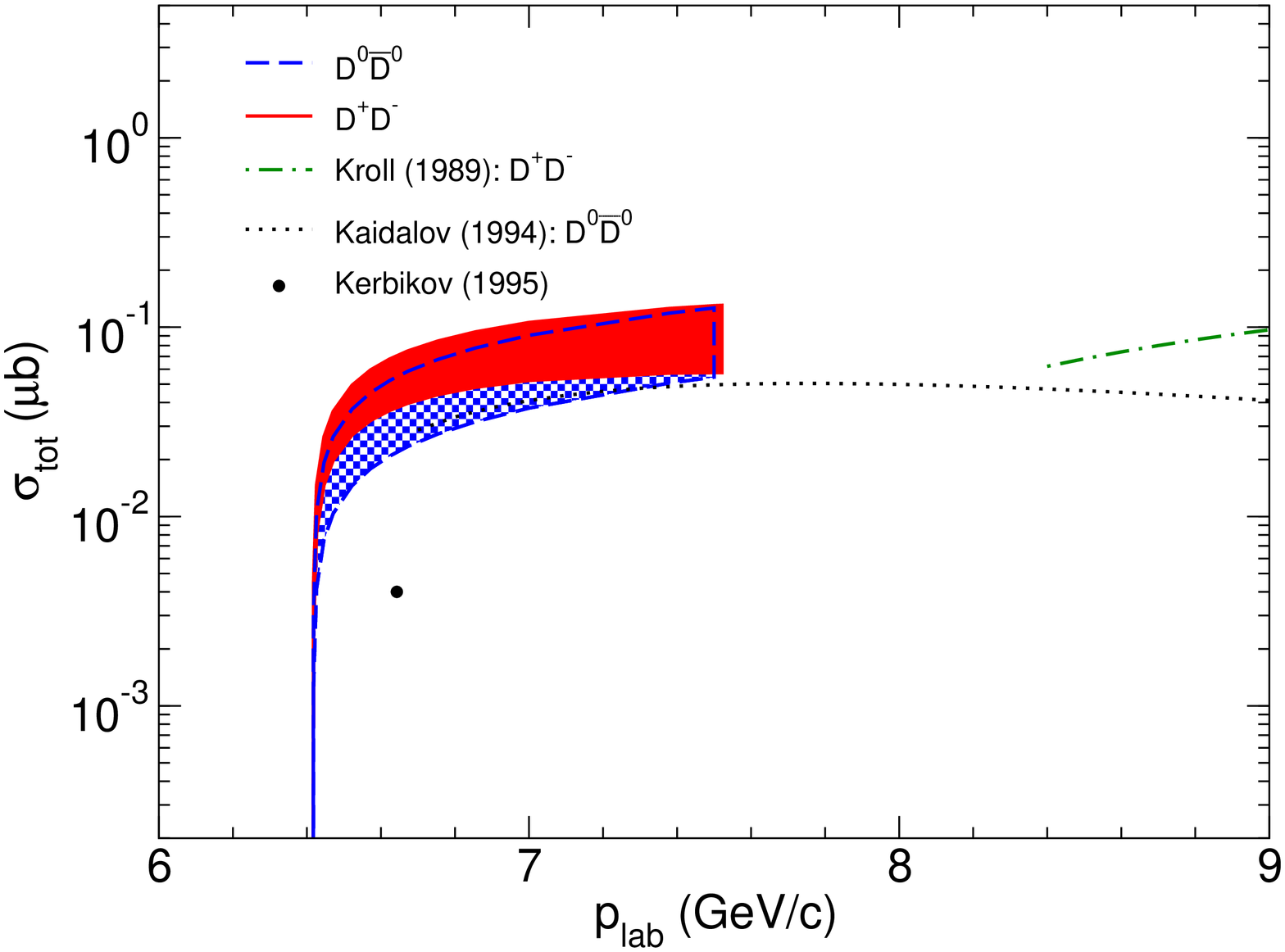}
\caption{Total reaction cross sections for
$\pbarp \to \ddbar$ as a function of $p_{lab}$. 
Left: Effects of the initial and final state interaction. 
Right: Our predictions (shaded band, grid) in comparison to those
of Refs. \cite{Kroll:1988cd}, \cite{Kaidalov:1994mda} and \cite{Kerbikov}. 
}
\label{fig:3}       
\end{figure}

Our predictions for the reaction $\pbarp \rightarrow \ddbar$ 
are presented in Fig.~\ref{fig:3}. Let us first focus on the
effects of the inital- and final state interaction. The
transition from $\pbarp$ to $\ddbar$ is generated by the
exchange of charmed baryons, in particular the $\Lambda_c$ and
$\Sigma_c$. Under the assumption of $SU(4)$ symmetry the
pertinent coupling constants are given by 
\begin{eqnarray}
f_{\Lambda_c N D} &=& 
-\frac{1}{\sqrt{3}}(1+2{\alpha}) f_{NN\pi} \approx {-1.04}\, f_{NN\pi}, 
\nonumber \\
f_{\Sigma_c N D} &=&
(1-2{\alpha}) f_{NN\pi} \approx {0.2}\, f_{NN\pi}, 
\nonumber
\end{eqnarray}
where we assumed for the F/(F+D) ratio ${\alpha} \approx 0.4$. 
Thus, one expects that $\Lambda_c$ exchange dominates the transition
while $\Sigma_c$ exchange should be suppressed. Specifically, this
implies that ${V}^{\pbarp \to {D^0 \bar D^0}} \gg {V}^{\pbarp \to {D^+ D^-}}$. 
Indeed, within the Born approximation the cross sections predicted 
for $D^0 \bar D^0$ are more than two orders of magnitude larger than
those for $D^+ D^-$ \cite{Titov:2008yf}, cf. the dotted lines Fig.~\ref{fig:3}. 
(The curves for $D^0 \bar D^0$ are marked with a circle.) 
However, once the ISI is included the situation changes drastically
and now both channels are produced at a comparable rate (dashed lines).
In fact, now the cross section for $D^+ D^-$ is even somewhat larger
than the one for $D^0 \bar D^0$. The inclusion of the FSI modifies
the predictions once again, though only on a quantitative level. 

On the right-hand side of Fig.~\ref{fig:3} we compare our predictions with 
those by other groups. 
Here our results are shown as a shaded band for $D^+ D^-$ and as grid 
for $D^0 \bar D^0$. Again the bands represent the variation of the 
predictions when different ISI's are used. Displayed are also
the predictions by Kroll et al. \cite{Kroll:1988cd} (dash-dotted line),
Kaidalov \cite{Kaidalov:1994mda} (dotted line) and by 
Kerbikov \cite{Kerbikov} (circle). Obviously, and unlike in the 
$\lcbarlc$ case, now the majority of the predictions are pretty much
comparable, at least on a qualitative level. 

\section{Summary}
\label{sec:4}
We have presented results for the reaction $\pbarp \to \lcbarlc$ 
of a model calculation performed within the meson-exchange picture 
in close analogy to the J\"ulich analysis of
the reaction $\pbarp \to \lbarl$ utilizing $SU(4)$ symmetry. 
The predicted cross sections are in the order of 1 -- 7 $\mu b$. Thus, 
they are about 10-100 times smaller than those for $\pbarp \to \lbarl$. 
However, and surprisingly, our predictions turned out to be about 1000 
times larger than those obtained in other model calculations. 

We presented also predictions for $\pbarp \to \ddbar$
obtained in the same spirit, i.e. by connecting this reaction with 
$\pbarp \to \bar K K$ via $SU(4)$ symmetry. 
Here the cross sections were found to be in the order of 
$10^{-2}$ -- $10^{-1}$ $\mu b$ and they turned out to be 
comparable to those predicted by other model calculations.

\begin{acknowledgements}
This work was partially financed by CNPq and FAPESP (Brazilian
agencies).
\end{acknowledgements}



\end{document}